\documentclass[preprint2]{aastex}
\usepackage{amssymb}
\usepackage{amsmath}
\usepackage{graphicx}
\usepackage{txfonts}

\shorttitle{Fast magnetoacoustic wave trains}
\shortauthors{Shestov, Nakariakov \& Kuzin}

\begin{document}
\title{Fast magnetoacoustic wave trains of sausage symmetry in cylindrical waveguides of the solar corona}


\author{S.~Shestov\altaffilmark{1} and V.~M.~Nakariakov\altaffilmark{2} and S.~Kuzin\altaffilmark{1}}
\altaffiltext{1}{Lebedev Physical Institute,  Leninskii prospekt, 53, Moscow, Russia, 119991}
\email{sshestov@gmail.com}
\altaffiltext{2}{Centre for Fusion, Space and Astrophysics, Department of Physics, University of Warwick, Coventry CV4 7AL, UK}
\date{Received ..; accepted ...}

\begin{abstract}
Fast magnetoacoustic waves guided along the magnetic field by plasma non-uniformities, in particular coronal loops, fibrils and plumes, are known to be highly dispersive, which leads to the formation of quasi-periodic wave trains excited by a broadband impulsive driver, e.g. a solar flare. We investigated effects of cylindrical geometry on the fast sausage wave train formation. We performed magnetohydrodynamic numerical simulations of fast magnetoacoustic perturbations of a sausage symmetry, propagating from a localised impulsive source along a field-aligned plasma cylinder with a smooth radial profile of the fast speed. The wave trains are found to have pronounced period modulation, with the longer instant period seen in the beginning of the wave train. The wave trains have also a pronounced amplitude modulation. Wavelet spectra of the wave trains have characteristic tadpole features, with the  broadband large-amplitude heads preceding low-amplitude quasi-monochromatic tails. The mean period of the wave train is about the transverse fast magnetoacoustic transit time across the cylinder. The mean parallel wavelength is about the diameter of the waveguiding plasma cylinder. Instant periods are longer than the sausage wave cutoff period. The wave train characteristics depend on the fast magnetoacoustic speed in both the internal and external media, and the smoothness of the transverse profile of the equilibrium quantities, and also the spatial size of the initial perturbation. If the initial perturbation is localised at the axis of the cylinder, the wave trains contain higher radial harmonics that have shorter periods.   
\end{abstract}


\section{Introduction}
Magnetohydrodynamic (MHD) waves with the periods from several seconds to
several minutes are ubiquitous in the corona of the Sun, and are
intensively studied in the contexts of heating and seismology of the coronal
plasma \citep[see, e.g.][for comprehensive recent reviews]{2012RSPTA.370.3193D,
2014SoPh..289.3233L}. Properties of MHD waves are strongly affected by
field-aligned structuring of the plasma, intrinsic for the corona
\citep[e.g.][]{2008A&A...491L...9V}. Filamentation of the coronal density along
the field leads to appearance of waveguides for magnetoacoustic waves
\citep[e.g.][]{zaj75, 1983SoPh...88..179E, 1984ApJ...279..857R}. In the
waveguides fast magnetoacoustic waves become dispersive, i.e. their phase and
group speeds become dependent on the frequency and wavelength. If the waves are
excited by a broadband, spatially localised driver, e.g. a flare or another
impulsive release of energy, induced fast magnetoacoustic perturbations
develop into a quasi-periodic wave train with pronounced amplitude and frequency
modulation \citep[][]{1984ApJ...279..857R,2004MNRAS.349..705N}. 

Possible manifestation of fast magnetoacoustic wave trains in solar
observations was first pointed out in quasi-periodic pulsations in solar radio
bursts detected at 303 and 343 MHz \citep[][]{1983Natur.305..688R,
1984ApJ...279..857R}. Later on, analysis of high-cadence imaging observations of the
green-line coronal emission recorded during a solar eclipse revealed the presence of
similar wave trains \citep{2002MNRAS.336..747W, 2003A&A...406..709K} with the
mean period of about 6 s. More recently, quasi-periodic wave trains with the
periods of several tens of minutes were detected in the decametric solar
emission \citep[see][]{2009ApJ...697L.108M, 2011A&A...529A..96K, 2013SoPh..283..473M}. Recently
discovered quasi-periodic rapidly-propagating wave trains of the EUV intensity,
with the periods about from one to a few minutes \citep{2011ApJ...736L..13L}
are possibly associated with this phenomenon too \citep{2013A&A...554A.144Y,2014A&A...569A..12N}. 

Identification of the physical mechanism for the formation and evolution of
coronal quasi-periodic wave trains requires advance theoretical modelling.  The
initial stage of the evolution of a broadband fast magnetoacoustic perturbation
in a quasi-periodic wave train was numerically simulated by
\citet{1993SoPh..144..101M, 1994SoPh..151..305M}. The mean period of the wave
train was found to be of the order of the fast wave travel time across the
waveguide, which was consistent with qualitative estimations by
\cite{1983Natur.305..688R}. 

\cite{2004MNRAS.349..705N} modelled the developed stage of the broadband fast
wave evolution in a plasma slab, and found that the resultant wave train has a
characteristic signature in the Morlet wavelet spectrum, the \lq\lq crazy
tadpole\rq\rq\ with an narrowband extended tail followed by a broadband \lq\lq
head\rq\rq. Similar wavelet spectral signatures are detected in observations
\citep{2009ApJ...697L.108M, 2011A&A...529A..96K, 2013A&A...554A.144Y}.  These
characteristic wavelet spectra appear in fast wave trains guided by plasma
slabs with the transverse profiles of the fast speed of different steepness.
However, it was demonstrated that the specific width of the wavelet spectral
peak and its time modulation are determined by the spectrum of the initial
perturbation  and the cutoff wavenumber \citep{2005SSRv..121..115N}. In
particular, in some cases the effect of dispersive evolution can lead to the
formation of quasi-monochromatic fast wave trains, without noticeable variation
of the period. Similar behaviour was established for fast wave trains guided by
plane current sheets \citep{2012A&A...537A..46J, 2014ApJ...788...44M}. Obvious
similarities of theoretical and observational properties of coronal fast wave
trains provide us with a tool for diagnostics of the waveguiding plasma
non-uniformities.

In addition, it was recently proposed that another common feature of broadband
solar radio emission, fiber bursts, could also be associated with guided fast
wave trains \citep{2013A&A...550A...1K}. Furthermore, it was demonstrated that
modulations of broadband radio emission produced in pulsars could also be
interpreted in terms of fast magnetoacoustic wave trains
\citep{2013A&A...552A..90K}.

More advanced models, while still in the plane geometry, accounting for the
curvature and longitudinal non-uniformity of the waveguide (or anti-waveguide)
showed that the formation of a quasi-periodic fast wave train is a robust
feature \citep{2013A&A...560A..97P, 2014A&A...568A..20P} that is
well-consistent with observations \citep{2014A&A...569A..12N}. Initial
perturbations of both sausage and kink symmetries were found to result in
quasi-periodic wave trains with the periods prescribed by the fast
magnetoacoustic transverse transit time in the vicinity of the initial perturbation.
Recent analytical modelling of the evolution of broadband fast waves in a
cylindrical waveguide with a step-function transverse profile, which represent,
e.g. coronal loops or prominence fibrils, demonstrated formation of wave trains too \citep{2014ApJ...789...48O, 2015ApJ...806...56O}.

The aim of this paper is to study the formation of fast wave trains in a
cylindrical waveguide with a smooth transverse profile, investigating the effect
of the cylindrical geometry and the waveguide steepness on the wavelet spectral
signatures of the wave trains. Our paper is structured as follows:
Section~\ref{numerical_setup} describes numerical setup, Section~\ref{results}
presents simulation results and their discussion. Finally,
Section~\ref{conclusions} gives the conclusions.

\section{Numerical setup \& Initial conditions}\label{numerical_setup}

The simulations were performed using the numerical code Lare3D \citep{2001JCoPh.171..151A}. This code
solves resistive MHD equations in the normalised Lagrangian form:

\begin{equation}
	\frac{D \rho}{Dt} = - \rho \nabla \cdot \mathbf{v},
\end{equation}

\begin{equation}
	\frac{D \mathbf{v}}{Dt} = \frac{1}{\rho} (\nabla \times \mathbf{B}) \times \mathbf{B} - \frac{1}{\rho} \nabla P,
\end{equation}

\begin{equation}
	\frac{D \mathbf{B}}{Dt} =  (\mathbf{B} \cdot \nabla) \mathbf{v}  -
	   \mathbf{B} (\nabla \cdot \mathbf{v}) - \nabla \times (\eta \nabla \times \mathbf{B} ),
\end{equation}

\begin{equation}
	\frac{D \epsilon}{Dt} = - \frac{P}{\rho} \nabla \cdot \mathbf{v} + \frac{\eta}{\rho} j^2,
\end{equation}
where $\rho$, $\epsilon$, $P$, $\mathbf{v}$, $\mathbf{B}$ and $\mathbf{j}$ are the mass density, specific internal
energy, thermal pressure, velocity, magnetic field and electric current density, respectively; $\eta$ is the electrical
resistivity.  Thermodynamical quantities are linked with each other by the state equation $P=\rho \epsilon (\gamma -1)$
where $\gamma$ is the ratio of specific heats. As in this study we are not interested in the dissipative processes, we
take $\gamma=5/3$.
The physical quantities were normalised with the use of the following constants: lengths are normalised to $L_0=1$~Mm,
magnetic fields to $B_0=20$~G and densities to $\rho_{00}=1.67 \times 10^{-12}$~kg~cm$^{-3}$.  The mass density
normalisation corresponds to the electron concentration $n_\mathrm{e0} = 10^9$~cm$^{-3}$. The normalising speed was
calculated as $v_0 = {B_0}/\sqrt{\mu_0 \rho_{00}} = 1,380$~km s$^{-1}$ that is the Alfv\'en speed corresponding to the
values $B_0$ and $\rho_{00}$.

The simulations were performed in a $160^3$-grid box, which corresponded to the physical volume of 4x4x20~Mm. The
resolution was uniform along the Cartesian axes, but different in the directions along and across the magnetic field,
allowing us to resolve fine scales in the transverse direction. 

Boundary conditions were set to be open (\textit{BC\_OPEN} in Lare3D), which is implemented via far-field Riemann
characteristics. The Lare3D authors stressed that the artificial reflection is typically a few percent but can be as
large as 10\% in extreme cases. {\bfseries Anyway, this reflection does not influence the results of our simulations, as our full attention is paid to the development of the pulse that freely propagates along the cylinder, before it reaches the cylinder's
end. In the online movie this perturbation propagates to the left. Hence, this pulse does not experience the reflection}. In order to test the undesirable effect of the wave reflection from the open $x$- and $y$-
boundaries we carried out test runs in a computational box with the doubled resolution, 240x240x160, and also increasing
the computational box in the transverse directions, 6x6x20~Mm. {\bfseries The correlation coefficient between the signals
corresponding to the developed wave trains amounted to $99.5$~\%, hence justifying the robustness of the results
obtained}.

The equilibrium plasma configuration was a plasma cylinder directed along the straight magnetic field, along the
$z$-axis. The cylindrical plasma non-uniformity was implemented by the radial profiles of the physical quantities
$B_z(r)$, $n_e(r)$, $T(r)$.  We considered cases of steep and smooth boundaries of the cylinder.  The case with the
step-function profiles (\textit{setup1}) was implemented with the following plasma parameters:  
$B_z=19.5$~G, $n_\mathrm{e}=3.3 \times 10^9$~cm$^{-3}$,  $T=9 \times 10^5$~K for $r \le R_0$; and
$B_z=20.0$~G, $n_\mathrm{e}=6.6 \times 10^8$~cm$^{-3}$,  $T=2 \times 10^5$~K for $r > R_0$ , where $R_0=1$~Mm is the cylinder radius.

\begin{figure}[pth]
	\begin{center}
		\includegraphics[width=8cm]{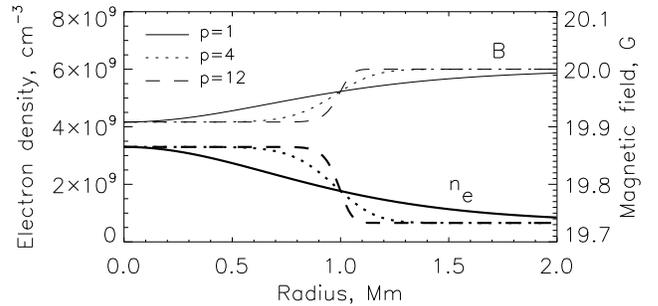}
	\end{center}
	\caption{Radial profiles of the electron concentration $n_\mathrm{e}(r)$ and magnetic field
	$B_z(r)$ in the simulated plasma cylinder. At the central axis the plasma concentration $n_e$ has a factor of 5 enhancement.}
	\label{fig:density_profile}
\end{figure}

The cases with smooth profiles (that are refereed to as \textit{setup2}, \textit{setup3}, \textit{setup4}, and \textit{setup5}) were implemented as follows: the temperature $T$ was constant throughout the whole volume, $n_\mathrm{e}(r)$ was given by the generalised symmetric Epstein function \citep{1995SoPh..159..399N, 2003A&A...409..325C}, and $B_z(r)$ was set to equalise the total pressure everywhere in the computational domain,
\begin{equation}
	\left\{ \begin{aligned}
            n_\mathrm{e}(r) & = n_{\infty}+ \frac{(n_0-n_{\infty})}{ \cosh^2 \left[ \left(r/R_0\right)^p \right] } \\
	    B_z(r)  & = B_{\infty} \left( 1 - \frac{ 16 \pi k_\mathrm{B} T (n_0 - n_{\infty})}{ B_{\infty}^2 \cosh^2 
	              \left[ \left(r/R_0\right)^p \right]} \right)^{1/2}, \\
    \end{aligned} \right. 
\end{equation}
where $T=2\times 10^5$~K, $B_{\infty}=20$~G, $n_{\infty}=6.6 \times 10^8$~cm$^{-3}$, $n_0=3.3 \times 10^9$~cm$^{-3}$,
$k_\mathrm{B}$ is the Boltzmann constant, and $p$ is the parameter controlling the boundary steepness. The specific
values of the parameter $p$ and other parameters of the equilibria, corresponding to different numerical setups, are
shown in Table~\ref{table1}. 
The radial profiles of the electron density $n_\mathrm{e}(r)$ and magnetic field $B_z(r)$ are depicted in
Fig.~\ref{fig:density_profile}.  We would like to point out that in all the setups the relative variation of the
magnetic field is small, and that the plasma-$\beta$ is much less than unity everywhere.

Dynamics of MHD waves is determined by the Alfv\'en, 
$v_\mathrm{A}(r)={B_z(r)}/\sqrt{\mu_0 \rho(r)}$, 
and sound
$C_\mathrm{S}(r)=\sqrt{\gamma P(r)/\rho(r)}$
speeds, where the mass density $\rho(r)$ is prescribed by the
electron concentration $n_\mathrm{e}(r)$, and the gas pressure $P(r)$ is determined by the mass density and the
temperature with the use of the state equation. In our simulations, the Alfv\'en speed was
$v_{\mathrm{A}\infty}=1,550$~km~s$^{-1}$ outside the cylinder, 
and $v_{\mathrm{A}0}=690$~km~s$^{-1}$ at the cylinder axis.
In \textit{setups 2--5} the sound speed was $C_\mathrm{S}=67.6$~km~s$^{-1}$ everywhere in the computational domain.  In
\textit{setups 1} the sound speed had the same value at the axis of the cylinder, and was
$C_\mathrm{S}=143.4$~km~s$^{-1}$ outside the cylinder. We would like to stress that in the low-$\beta$ plasma that is
typical for coronal active regions, the specific value of the sound speed does not affect the parameters of fast
magnetoacoustic waves, as their dynamics is controlled by the Alfv\'en speed.

In all initial setups the fast magnetoacoustic speed determined as $\sqrt{v_\mathrm{A}^2(r) + C_\mathrm{S}^2(r)}$ had a minimum at the axis of the cylinder. Thus, the plasma cylinder was a refractive waveguide for fast magnetoacoustic waves. 

The initial perturbation was set up as a perturbation of the \textbf{transverse} velocity of the sausage symmetry, 
\begin{equation}
	v_r = A r  \exp \left(-\frac{r^2}{2\sigma_r^2} \right)  \exp \left( -\frac{(z-z_0)^2}{2\sigma_z^2} \right) ,
\end{equation} 
where $\sigma_r=0.9$~Mm for \textit{setup 1}, \textit{setup 2}, \textit{setup 3}, \textit{setup 4} and
$\sigma_r=0.2$~Mm for the \textit{setup5} (c.f. with the unperturbed radius of the cylinder, $R_0=1$~Mm), and
$\sigma_z=0.2$~Mm and $A=0.1v_0$ for all the setups. The latter condition insures that nonlinear effects are negligible.
The pulse is axisymmetric, and hence generates sausage mode perturbations. We placed the pulse near one of the
boundaries in the $z$-direction, in order to give more space for the wave train to evolve when it propagates towards the
opposite end of the cylinder. 

\begin{deluxetable}{llccccc}
	\tablecaption{Equilibrium conditions 
	}
	\tablehead{
	  \colhead{Title} & \colhead{Profile} & \colhead{$B_\infty$/$B_0$, G} &
	  \colhead{$n_\infty$/$n_0$, $\times 10^9$ cm$^{-3}$} &
	  \colhead{$T_\infty$/$T_0$, $\times 10^5$ K} &	\colhead{$v_{A_\infty}$/$v_{A_0}$, km s$^{-1}$} &
	  \colhead{$\sigma_r$, Mm}
	}
	\startdata
	\textbf{setup1} & step-function, $p=\infty$ & 20.0/19.5 & 0.66/3.3 & 9.0/2.0 &   $1 550$/680 & 0.9 \\
	\textbf{setup2} & smooth-profile, $p=12$ & 20.0/19.9 & 0.66/3.3 & 2.0/2.0 &   $1 550$/690 & 0.9 \\
	\textbf{setup3} & smooth-profile, $p=4$  & 20.0/19.9 & 0.66/3.3 & 2.0/2.0 &   $1 550$/690 & 0.9 \\
	\textbf{setup4} & smooth-profile, $p=1$  & 20.0/19.9 & 0.66/3.3 & 2.0/2.0 &   $1 550$/690 & 0.9 \\
	\textbf{setup5} & smooth-profile, $p=4$  & 20.0/19.9 & 0.66/3.3 & 2.0/2.0 &   $1 550$/690 & 0.2\\
	\enddata
	\tablecomments{Parameters of the initial equilibrium in different numerical runs.}
\label{table1}	
\end{deluxetable}

\section{Results and discussion}\label{results}

\begin{figure*}[pth]
	\begin{center}
		\includegraphics[width=15.5cm]{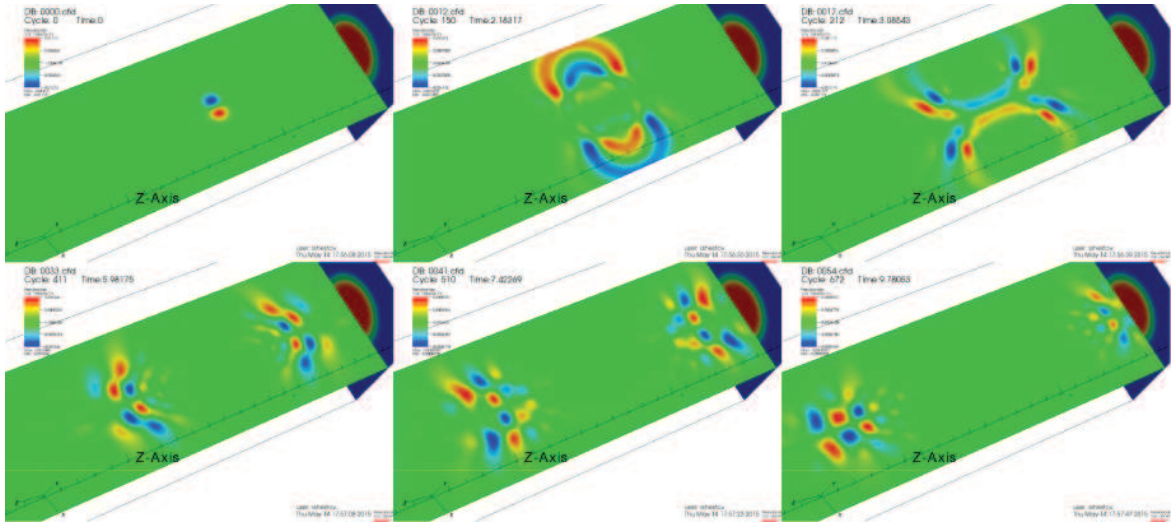}
	\end{center}
	\caption{Initial perturbation of the sausage symmetry, corresponding to
	\textit{setup5} (the upper panels), and the development (lower panels)
	of fast magnetoacoustic wave trains in the magnetic tube. The snapshots
	correspond to the times $t=$0, 2.18, 3.09, 5.98, 7.42, 9.18~s elapsed
	after the initial excitation. The horizontal slice, parallel to the
	cylinder axis along the field, in the $z$-direction, denotes the \textbf{transverse
	flow}; the perpendicular slice denotes the \textbf{equilibrium density profile}.
        An animation and a color version of this figure are available in the online journal.}
	\label{fig:setup4-3-3D}
\end{figure*}

Figure~\ref{fig:setup4-3-3D} shows evolution of a compressive pulse in the plasma cylinder.  The initial perturbation
develops in a compressive wave that initially propagates radially and obliquely to the axis of the cylinder (\textbf{see
the top central panel}), gets partially reflected on gradients of the equilibrium physical parameters, and returns back
to the axis of the cylinder (\textbf{see the top right panel}).  Then the perturbation overshoots the axis, gets
reflected from the opposite boundary, and this scenario repeats again and again (\textbf{see the bottom panels}). Thus,
the fast magnetoacoustic perturbation is guided along the axis of the field-aligned plasma cylinder. {\bfseries This
phenomenon is well-known in the solar literature since the pioneering works of \citet{zaj75} and
\citet{1983SoPh...88..179E}.  In more detail the formation of the wave train can be watched in the online movie.
The basic physics of this phenomenon is connected with the geometrical dispersion: waves
with different wavelengths propagate at different phase and group speeds. Thus, for the same time after the
excitation at the same point in the waveguide, different spectral components of the initial excitation travel different distances, and the
perturbation becomes \lq\lq diffused\rq\rq\ along the waveguide. Hence, after some time an initially broadband
perturbation develops in a quasi-periodic wave train with frequency and amplitude modulation \citep[see,
e.g.][]{2004MNRAS.349..705N}}. 

The geometrical dispersion caused by the presence of a characteristic spatial scale in
the system, the diameter of the cylinder, leads to the formation of two quasi-periodic fast wave trains propagating in
the opposite directions along the axis of the waveguide. This finding is consistent with the analytical results of
\citet{2014ApJ...789...48O, 2015ApJ...806...56O}.  In the trains, the characteristic parallel wavelengths are comparable
by the order of magnitude to the diameter of the waveguide. {\bfseries In the Cartesian coordinates used in the figure,
the transverse plasma flows induced by the perturbations are odd functions with respect to the transverse coordinate
with the origin at the cylinder's axis. The same structure of the transverse flows is seen in any plane including the
axis of the cylinder. In other words the perturbations are independent of the azimuthal angle in the cylindrical
coordinates with the axis coinciding with the axis of the cylinder. The transverse flows vanish to zero at the axis of
the cylinder. Thus the excited perturbations are of the sausage symmetry, prescribed by the symmetry of the initial
perturbation}. 

\begin{figure*}[pth]
	\begin{center}
		\includegraphics[width=16.0cm]{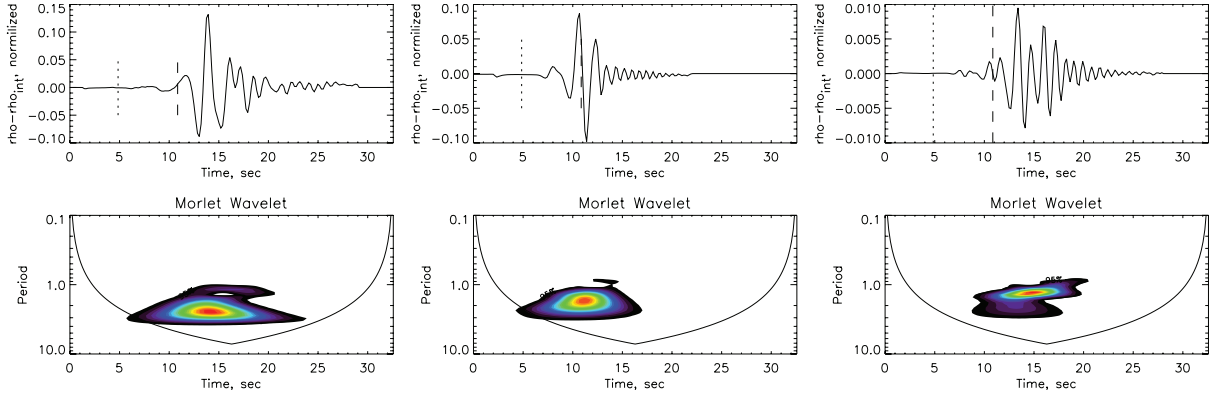}
	\end{center}
	\caption{Numerical simulation of an impulsively generated fast
	magnetoacoustic wave train propagating along a field-aligned plasma cylinder for different numerical setups:
	left --- a sharp (step-function) density profile with $p=\infty$ (\textit{setup1}), middle --- smooth density profile with $p=1$
	(\textit{setup4}), and right --- smooth density profile with $p=4$ but with  narrower initial perturbation
	radial size $\sigma_r=0.2$ (\textit{setup5}). Upper panels: the time evolution of the perturbation of density
	$\rho(t)-\rho_0$ at the observational point situated at the
	magnetic tube axis at $h = 7.5$~Mm from the location of the initial perturbation. The vertical
	lines show the pulse arrival time if the density was uniform: the dotted
	line corresponds to the fast travel time calculated using the external density $\rho_{\infty}$, and the dashed line was obtained with the use of the internal
	density $\rho_0$. Lower panels: Morlet wavelet transform of the density variation signal at the observational point, demonstrating the
	characteristic \lq\lq tadpole\rq\rq\ wavelet signature.}
	\label{fig:profiles_merged}
\end{figure*}

Characteristic time signatures of the developed fast magnetoacoustic wave train formed in the waveguides with different
equilibrium profiles \textit{setup1}, \textit{setup4}, and \textit{setup5}, are shown in Fig.~\ref{fig:profiles_merged}.
{\bfseries In this study we considered only the \lq\lq direct\rq\rq\ wave trains which propagated along the cylinder
before they reached its end and experienced reflection (in Fig.~\ref{fig:setup4-3-3D} and the online movie it is the wave train
propagating to the left)}.  The upper panels show time evolution of the density measured at the axis at the distance
$h=7.5$~Mm from the initial perturbation location ($z_0$).  The lower panels show the Morlet wavelet spectra of the
signals. All the signals are seen to have similar features typical for guided fast wave trains: frequency and amplitude
modulations, characteristic and cutoff periods, the arrival time determined by $v_{A\infty}$, and the \lq\lq
average\rq\rq\ speed determined by $\sim v_{A0}$, dependence of their temporal characteristics on the radial profile of
the waveguiding plasma non-uniformity and the spatial size of the initial perturbation. All the temporal characteristics
of the wave trains formed in the \textit{setup2} and \textit{setup3}, with $p=12$ and $p=4$, respectively, are similar
to those of the \textit{setup1} (the step-function profile) and are not shown here.

The fast wave trains are seen to have a characteristics period of 2--3~s, which is in an agreement with the estimation
provided by \citet{1984ApJ...279..857R}:
\begin{equation}
        P_\mathrm{prop} \approx \frac{2\pi R_0}{j_0 v_{\mathrm{A}0}} \sqrt{1-\frac{\rho_\infty}{\rho_0}} \approx 3.4 \mbox{~s},
\end{equation}
where $j_0 \approx 2.40$ is the first zero of the Bessel function $J_0(x)$. 

Sausage modes of a low-$\beta$ plasma cylinder experience a long-wavelength cutoff \citep{1983SoPh...88..179E}.  The
spectral components with wavelengths longer than the cutoff value (or with the periods longer than the cutoff period)
are not trapped in the cylinder, and leak out \citep[see, e.g.][]{1986SoPh..103..277C, 2012ApJ...761..134N,
2013A&A...560A..97P}.  Thus, these spectral components are not present in the spectrum of the guided wave trains. In the
case of the step-function profile the cutoff period is
\begin{equation}
	P_\mathrm{cutoff}=\frac{\sqrt{2} \pi R_0 }{v_{{\mathrm{A}0}}} \sqrt{ 1-\frac{\rho_{\infty}}{\rho_0} } \approx 5.7 
	\mbox{~s},
\end{equation}
\citep[e.g.][]{1983SoPh...88..179E}. According to the wavelet analysis
(Fig.~\ref{fig:profiles_merged}) the observed maximum period is about 3.5~s, which is roughly
consistent with the estimation. The small discrepancy between the theoretical estimation and the numerical result can be
attributed to the wave train nature of the signal. Indeed, in the signal the longest period seen in the very beginning
of the train lasts for about one cycle of the oscillation only, which leads to effective broadening of the spectrum.

Numerical simulations of the discussed phenomenon, performed for a plasma slab \citet{2004MNRAS.349..705N} and a plane
current sheet \citet{2012A&A...537A..46J} showed that the wavelet spectra of a developed wave train have a
characteristic \lq\lq crazy tadpole\rq\rq\ shape: a long, almost monochromatic tail of low amplitude is followed by a
broad-band and high-amplitude head (as the tadpole goes tail-first, it was refereed to as \lq\lq crazy\rq\rq).  Another
characteristic feature of fast wave trains formed in waveguides is a pronounced period modulation: the instant period
decreases in time. According to the reasoning provided by \citet{1984ApJ...279..857R} this effect is connected with the
geometrical dispersion: longer-wavelength (long period) spectral components propagate at the higher speed, about
$v_{A\infty}$, and hence reach the observational point earlier.  In contrast, shorter-wavelength (shorter period)
components excited simultaneously and at the same initial location propagate at lower speeds that gradually decrease
from the value $v_\mathrm{A0}$ with the decrease in the wavelength (period) and hence reach the observational point
later.

Likewise, the wavelet spectra obtained for cylindrical waveguides, shown in Fig.~\ref{fig:profiles_merged} demonstrate
the period modulation with the decrease in the instant period with time. The modulation is even better seen in the time
signals. As for plane waveguides, wavelet spectra of fast wave trains guided by cylinders have tadpole features too.
However, the spectral features have either almost-symmetrical head-tail shapes, or the shapes with a broadband \lq\lq
head\rq\rq\ occurring in the beginning of the tadpole.  Indeed, the time signals of the wave trains show that in the
considered case the highest amplitude is reached not near the end of the train, as it is in the slab case, but near its
beginning. Thus, one can say that in this case the tadpole goes \lq\lq head-first\rq\rq, so they are not \lq\lq
crazy\rq\rq. The head of the wave train is seen to propagate at the speed about or slower than the Alfv\'en speed at the
axis of the waveguide, in agreement with the prediction made in \citet{1984ApJ...279..857R}. Thus, we obtain that the
characteristic wavelet spectral shapes of fast wave trains are different in the cases of a slab and cylinder waveguides.
In both the geometries the instant period decreases with time, while the amplitude modulation of the trains is different
in these two cases. It provides us with a potential tool for the observational discrimination between these two plasma
non-uniformities in frames of MHD seismology. 

Fast wave trains formed in plasma cylinders with smooth radial profiles (with $p<\infty$, \textit{setups 2--5}) are seen
to have similar characteristic modulations of the instant amplitude and period. It confirms the robustness of this
effect, as it is not very sensitive to the specific shape of the radial profile. Unfortunately, in the cylindric case we
are not aware of the existence of an exact analytical solution of the eigen value problem for trapped fast
magnetoacoustic waves, similar to the solution described in \citep{1995SoPh..159..399N, 2003A&A...409..325C} for the
symmetric Epstein transverse profile. 

Another interesting feature of the modelling is the appearance of \lq\lq fins\rq\rq\ in the tadpole spectral features
(see left panel in Fig.~\ref{fig:profiles_merged}; similar features are also present in \textit{setup2} and \textit{setup3}, not
showed here).  The fins are associated with the excitation of the higher radial harmonics. Indeed, it is more pronounced
in the case of \textit{setup5}, when the initial pulse has a short size in the radial direction in comparison with the
cylinder's radius, $\sigma_z=0.2 \ll R_0 = 1$.  Thus, in this case the spatial radial spectrum of the initial
perturbation is broader and less close to the fundamental radial harmonic, leading to the effective excitation of the
higher radial harmonics \citep{1995SoPh..159..399N,2014ApJ...788...44M}. The appearance of the \lq\lq fins\rq\rq\ can be
readily understood from the dispersion plot \citep[see, e.g.][]{1983SoPh...88..179E}: for the same arrival time, and
hence the same group speeds, higher radial harmonics have shorter wavelengths (shorter periods) than the fundamental
radial harmonics. This feature is robust and is not qualitatively modified in the case of a smooth radial profile.

\section{Conclusions}\label{conclusions}

We summarise our findings as follows:
\begin{itemize}
	\item Dispersive evolution of impulsively generated fast
		magnetoacoustic waves of sausage symmetry, guided by a plasma cylinder representing e.g. a coronal loop, a filament fibril, a polar plume, etc., leads to the 
		development of quasi-periodic wave trains. The trains are similar to that found in the plane geometry and also recently analytically. 
		The wave trains have a pronounced period modulation, with the longer instant period seen in the beginning of the wave train. The wave trains have also a pronounced amplitude modulation. The main
		characteristics of the developed wave trains conform mainly 
		with the theoretically and numerically predicted. Wavelet spectra of the wave trains have characteristic tadpole features, with the \lq\lq heads\rq\rq\ preceding the \lq\lq tails\rq\rq. Similar features have been found in the plane case, while there the tadpoles typically go \lq\lq tail-first\rq\rq.
	\item The mean period of the wave train is about the transverse fast magnetoacoustic transit time across the cylinder.
	\item The mean parallel wavelength is about the diameter of the waveguiding plasma cylinder.
	\item Instant periods are longer than the cutoff period.	
	\item The wave train characteristics depend on the fast magnetoacoustic speed in both the internal and external media, and the smoothness of the transverse profile of the equilibrium quantities, and also the spatial size of the initial perturbation. 	
	\item In the case of the initial perturbation localised at the axis of the cylinder, the wave trains contain not only the fundamental radial harmonics, but also higher radial harmonics that have shorter periods, but arrive at the remote observational point simultaneously with the fundamental radial harmonic.
\end{itemize}

We conclude that fast magnetoacoustic wave trains observed in the solar corona in the radio, optical and EUV bands are a good potential tool for the diagnostics of the waveguiding plasma structures. The main problem in the implementation of this technique is the need for the information about the initial excitation, that possibly could be excluded in the case of observation of the wave train at different locations. This problem is under investigation and will be published elsewhere.

The work was partially supported by the Russian Foundation of Basic Research (grants 13-02-00044 and 14-02-00945). VMN acknowledges the support by the European Research Council  Advanced Fellowship No.~321141 \textit{SeismoSun}.

\bibliographystyle{hapj}   
\bibliography{shestov_vmn}

\end{document}